\documentclass[a4paper]{jpconf}
\usepackage{graphicx}
\begin{document}
\title{Radiation Hydrodynamics Scaling Laws in High Energy Density Physics and Laboratory Astrophysics}

\author{\'Emeric Falize$^{1,2}$, Serge Bouquet$^{1,2}$, Claire Michaut$^{2}$}

\address{$^{1}$ D\'epartement de Physique Th\'eorique et Appliqu\'ee, CEA/DIF, 91680 Bruy\`eres-le-Ch\^atel, France} 
\address{$^{2}$ LUTH, Observatoire de Paris, CNRS, Universit\'e Paris Diderot ; 5 Place Jules Janssen, 92190 Meudon, France}

\ead{emeric.falize@cea.fr}
\begin{abstract}
In this paper, radiating fluids scaling laws are studied. We focus on optically thin and optically thick regimes 
which are relevant for both astrophysics and laboratory experiments. By using homothetic Lie groups, we obtain the scaling laws, the similarity properties and the number of free parameters which allow to rescale experiments in the two astrophyscial situations. 
\end{abstract}

\section{Introduction}
High-Energy-Density Physics is a new way for astrophysicists to explore phenomena usually occurring 
in the Universe. The use of powerful facilities, enables us to bring the matter up to extreme states of density
and temperature in laboratory \cite{Remington06}. The astrophysical relevance of these experiments can be 
checked from scaling laws provided the physical system under study satisfies similarity properties. Thus, scaling laws and similarity properties must be examined with rigorous formalism. Several studies have been published about similarity and scaling laws.
 For instance, in \cite{Ryutov99}, \cite{Ryutov00}, purely hydrodynamics and MHD scaling laws are respectively considered and in \cite{Ryutov99}, the Birkhoff
  polytropic symmetries \cite{Birkhoff55} are recovered. Moreover, in \cite{Ryutov01} optically thin radiative hydrodynamic scaling laws have been considered 
  and in \cite{Castor07}, the author  not only studied similarity in case of optically thin plasma too but, also, discussed non-LTE situations through a microscopic approach. All these works have been carried out in an astrophysical context and were mainly based upon dimensional arguments. Scaling laws were also obtained for the Inertial Confinement Fusion (ICF) \cite{Murakami02}, \cite{Levedahl97} in order to determine the minimum energy required for ignition. These are very interesting too because they can be used as non trivial tests for numerical simulations. In this paper, we study the radiating fluid similarity problem in two 
  different regimes that can be (or will be) achieved in laboratory with current or future facilities. In each case, we derive the corresponding scaling laws 
  and in order to get rigorous and exact relations, our approach is based on the Lie groups \cite{Ibragimov99}. In the first part, we describe this method and 
  remind its fundamental concepts. The second part deals with the optically thin radiating fluids, which are a major topic in astrophysics. Comparisons with other
   results obtained earlier are carried out. Finally we consider the equilibrium diffusion approximation including radiative pressure and energy. For each approximation, 
   connections with astrophysical objects are provided and we emphasize the number of free parameters left to rescale an experiment.  
\section{Lie groups, similarity and scaling laws}
The invariant transformation group theory elaborated by Sophus Lie is a very powerful tool of theoritical physics to study the symmetry properties of partial differential equations (PDE) and to perform their analytical integration. Among all Lie groups, one of them, namely, the one-parameter homothetic group (HG) is frequently used, first because of its simplicity and, then, because it provides more general self-similar solutions 
  than those derived from dimensional analysis. This property  arises because the HG is a sub-group of scaling transformations. Now, remembering the philosophy of 
  Laboratory Astrophysics (\emph{i.e.} to recreate  systems having astronomical size on short scales), it seems natural to use the HG in order to study similarity 
  properties, scaling laws and even self-similarity. Here, we will focus on the first two points only. Group invariance of PDE together with 
  their solutions implies that the initial conditions (IC) be preserved from the laboratory system to the astrophysical one. This intuitive but constraining condition 
  is discussed in details in \cite{Ryutov99}, \cite{Ryutov00}. From this property, experimental data and IC in laboratory provide, first, information about astrophysical
   environments and, second, a transposition to astronomical objects. Moreover, the invariance of equations by HG implies that the Rankine-Hugoniot relations are 
   also invariant and, therefore, we make sure that small scale shocks correspond to the homothetic structures of astrophysical shocks.

\section{Similarity and scaling laws of optically thin radiating fluids}
When the cooling (or heating) characteristic time of a plasma gets close to its dynamical time, this should be considered in the modeling. Concerning optically thin plasmas, \emph{i.e.} $\lambda_{p}>>L$ ($\lambda_{p}$ is the mean free path of photons and $L$ is the 
characteristic plasmas of the system), a simple modeling of radiating losses (or heating) can be done simply by introducing a loss (or gain) of entropy. Thus, the plasma is described by the following equations:
\begin{equation}\label{eq1:eq}
\frac{\partial \rho}{\partial t}+\vec{\nabla}.[\rho \vec{v}]=0,\quad \rho\frac{d\vec{v}}{dt}=-\vec{\nabla}P_{th},\quad \frac{d P_{th}}{dt}-\gamma\frac{P_{th}}{\rho}\frac{d\rho}{dt}=-(\gamma-1)\mathcal{L}(\rho,T),\quad dM=\rho.dV,
\end{equation}
where $d/dt$ is the Lagrangian derivative and $\rho$, $\vec{v}$, $P_{th}$, $\gamma$ and $M$ are respectively the density, velocity, thermal pressure, polytropic 
index and the mass of the fluid. The function $\mathcal{L}(\rho,T)$ writes $\mathcal{L}(\rho,T) = \mathcal{Q}_{1}(\rho,T)+\mathcal{Q}_{2}(\rho,T)$
where $\mathcal{Q}_{1}$ and $\mathcal{Q}_{2}$ are energy sources (or losses). In addition, we assumed a 
polytropic evolution; \emph{i.e.} $P_{th}=(\gamma-1)\rho e$ where $e$ is the specific internal energy. 
Finally, an equation of state should be added to close (\ref{eq1:eq}): $P_{th}=\varepsilon_{0}[Z]\rho^{\mu}T^{\nu}$ where $\varepsilon_{0}[Z]$ is a function of the ionization 
Z. It should be noticed that to satisfy the first thermodynamical principle we should have $\gamma(1-\nu)=(\mu-\nu)$. Experimentally, heating can represent the laser energy 
deposition. From an astrophysical viewpoint, this modeling describes interstellar jets, bow shocks, radiating shocks (point C in Drake diagram \cite{Drake06}, Fig 7.17) in Polars and supernova remnants. The relation between the typical quantities in astrophysical objects and laboratory experiments (that we note with $\sim$) are 
given by:  $ r=a^{\delta_{1}}\tilde{r}$, $t=a^{\delta_{2}}\tilde{t}$, $\vec{v}=a^{\delta_{3}}\tilde{\vec{v}}$, $M=a^{\delta_{4}}\tilde{M}$, $\rho=a^{\delta_{5}}\tilde{\rho}$, $P_{th}=a^{\delta_{6}}\tilde{P}_{th}$,
 $\mathcal{Q}_{1}=a^{\delta_{7}}\tilde{Q}_{1}$, $\mathcal{Q}_{2}=a^{\delta_{8}}\tilde{Q}_{2}$, $\varepsilon_{0}=a^{\delta_{9}}\tilde{\varepsilon}_{0}$, $T=a^{\delta_{10}}\tilde{T}$, $\gamma=a^{\delta_{11}}\tilde{\gamma}$ 
 where $a$ is the group parameter and $\delta_{i}$ are the homothetic exponants. Rescaling $\varepsilon_{0}$, and $\mathcal{Q}_{i}$ can absorb a modification of Z from one system to the other (for example 
 in bremsstrahlung cooling $\mathcal{Q}\propto Z^{2}$ ).
  Up to now, the sources have not been specified but in the applications we will consider power law forms ($\mathcal{Q}_{i}=\mathcal{Q}_{0,i}\rho^{m_{i}}P^{n_{i}} r^{l_{i}}$). This type of source 
  is quite suitable for cooling since several processes in the continuum write in this simple form. We can also write $\mathcal{Q}_{i} \propto \kappa_{P}\sigma T^{4}$ where $\sigma$ is the Stefan-Boltzmann 
  constant and $\kappa_{P}$ is the Planck mean opacity that can be modeled by a power law at high temperature. The invariance of equations under the HG provides the 
  group invariants \cite{Ibragimov99} namely: $I_{1}=vt/r=St$ (Strouhal number), $I_{2}=\gamma$,  $I_{3}=P_{th}t/\rho v r=Eu\times St=St/[\gamma\mathcal{M}^{2}]$ ($Eu$: Euler number, $\mathcal{M}$: Mach number),
   $I_{4}=\mathcal{Q}_{1} t/P_{th}\propto t/t_{\mathcal{Q}_{1}}$, $I_{5}=\mathcal{Q}_{2} t/P_{th}\propto t/t_{\mathcal{Q}_{2}}$, where $t_{\mathcal{Q}_{i}}$ is the characteristic time of the sources $Q_{i}$ and  $I_{6}=M/[\rho r^{1+d}]$ (mass conservation).
\begin{table}
\caption{ \small{\emph{Scaling for optically thin plasmas for power law models of sources (second column). Plane (d=0) radiative shock problem for magnetic cataclysmic variables: 
the third column corresponds to Bremsstrahlung Cooling (BC) [$\Lambda\propto \rho^{2}T^{1/2}$] which can be Chevalier-Imamura unstable \cite{Chevalier82} and the fourth column is obtained for BC plus cyclotronic cooling (CC) [$\Lambda \propto \rho^{0.15} T^{2.5}$] and $\alpha=P_{th}\rho^{-\gamma}$.}}}
\begin{center}
\begin{tabular}{lllll}
\br
 \small{physical ratio} & \small{ratio} (scaling factor)  & BC & BC + CC \\
\mr
$r/\tilde{r}$ & $a^{\delta_{1}}$   & $a^{\delta_{6}-2\delta_{5}}$ & $a^{-3\delta_{5}/40}$ \\
$\rho/\tilde{\rho}$ & $a^{\delta_{5}}$    & $a^{\delta_{5}}$  & $a^{\delta_{5}}$  \\
$P/\tilde{P}$ & $a^{\delta_{6}}$    &$a^{\delta_{6}}$ & $a^{77\delta_{5}/40}$   \\
$t/\tilde{t}$& $a^{\delta_{1}+(\delta_{5}-\delta_{6})/2}$     & $a^{(\delta_{6}-3\delta_{5})/2}$ & $a^{-43\delta_{5}/80}$ \\
$v/\tilde{v}$ & $a^{(\delta_{6}-\delta_{5})/2}$    & $a^{(\delta_{6}-\delta_{5})/2}$  & $a^{37\delta_{5}/80}$  \\
$T/\tilde{T}$ & $a^{(\delta_{6}-\delta_{9}-\mu \delta_{5})/\nu}$   & $a^{(\delta_{6}-\delta_{5})}$ &  $a^{37\delta_{5}/40}$  \\
$M/\tilde{M}$ & $a^{\delta_{5}+(1+d)\delta_{1}}$  & $a^{\delta_{6}-\delta_{5}}$ & $a^{37\delta_{5}/40}$ \\
$\alpha/\tilde{\alpha}$ & $a^{\delta_{6}-\gamma \delta_{5}}$  & $a^{\delta_{6}-\gamma \delta_{5}}$ & $a^{(77/40-\gamma )\delta_{5}}$ \\
$\mathcal{Q}_{0,1}/\tilde{\mathcal{Q}}_{0,1}$ & $a^{(3/2-n_{1})\delta_{6}-(m_{1}+1/2)\delta_{5}-(l_{1}+1)\delta_{1}}$& 1 & 1 \\
$\mathcal{Q}_{0,2}/\tilde{\mathcal{Q}}_{0,2}$ & $a^{(3/2-n_{2})\delta_{6}-(m_{2}+1/2)\delta_{5}-(l_{2}+1)\delta_{1}}$ &  0 & 1  \\
\br
\end{tabular}
\end{center}
\end{table}
As expected, the invariants of this group are identical to the dimensionless numbers derived in similarity studies \cite{Castor07}. However, our approach is more general since we have local dimensionless quantities in contrast to global dimensionless numbers obtained thanks to the dimensional analysis. Thus, in our extension, the physical fields are conserved. Table 1 shows scaling laws for polars. Generally, we have four 
free parameters ($\delta_{1}$, $\delta_{5}$, $\delta_{6}$ and $\delta_{9}$) and, if we preserve ionization, only two (resp. one) exponent(s) remain(s) for a 
single source (resp. two sources). Moreover, if we set $\delta_{5}=0$, $\delta_{1}=1$, $\delta_{6}=2$ and $\mathcal{Q}_{2}=0$, we get the similarity considerations 
of \cite{Boily95}.  Thus, with the same formalism, we can study similarity properties, scaling laws, and include the specific case presented in \cite{Boily95}.  

\section{Similarity and scaling laws of optically thick radiating fluids}
Many systems, as well in laboratory as in astrophysics, are optically thick to radiation. For instance, the many classes of stars are more or less affected by radiation. Radiative pressure implies that there is an upper limit to the mass of a star (Eddington limit). Generally, including radiative flux in laboratory experiments is enough and that is why, 
researches about scaling laws in this regime have been carried out in ICF. Here, we add the energy and pressure of radiation ( see \cite{Mihalas99}) in the
 diffusion approximation at ETL. In experiments, LTE is usually satisfied \cite{Castor07} and it will be achieved on LMJ and NIF. In Astrophysics, radiation pressure and energy 
 play a key role in stars, supernovae, in evaporation phenomena, in clumps \cite{Konigl84}... The plasma evolution is then governed by the equations (\cite{Drake06}, pp 270-271):
\begin{equation}\label{eq5:eq}
\rho\frac{d\vec{v}}{dt}=-\vec{\nabla}[P_{th}+P_{rad}], \quad \frac{d}{dt}(\rho e+E_{rad})-\frac{\rho e+P_{th}+E_{rad}+P_{rad}}{\rho}\frac{d\rho}{dt}=-\vec{\nabla}.\vec{F}_{rad}-\mathcal{Q},
\end{equation}
where $\vec{F}_{rad}$, $E_{rad}$, $P_{rad}$ and $\mathcal{Q}$ are respectively the radiative flux, radiative energy density, radiative pressure and the energy source term.
 In the application, we will consider that $E_{rad}=a_{R}T^{4}$, $P_{rad}=E_{rad}/3$, $\vec{F}_{rad}=-\kappa_{rad}\vec{\nabla}T$ where $\kappa_{rad}$ is the radiative conductivity 
 given by $\kappa_{rad}=\kappa_{0}\rho^{m}T^{n}$[we still have $P_{th}=\varepsilon_{0}\rho^{\mu}T^{\nu}$]. In addition to the optically thin case we add the radiative 
 relations: $\vec{F}_{rad}=a^{\delta_{12}}\tilde{\vec{F}_{rad}}$; $\kappa_{rad}=a^{\delta_{13}}\tilde{\kappa}_{rad}$; $\kappa_{0}=a^{\delta_{14}}\tilde{\kappa}_{0}$;
  $ E_{rad}=a^{\delta_{15}}\tilde{E}_{rad}; P_{rad}=a^{\delta_{16}}\tilde{P}_{rad}; \mathcal{Q}=a^{\delta_{17}}\tilde{\mathcal{Q}}$. As before $I_{1}$, $I_{2}$, $I_{3}$, $I_{4}$ (or $I_{5}$) and $I_{6}$ 
  are five invariants. The additional ones are $I_{7}=P_{rad}t/(\rho v r)=Eu_{rad}\times St$ ($Eu_{rad}$: Radiative Euler number), $I_{8}=E_{rad}/P_{th}\propto 1/R$ ($R$: Mihalas numbers), 
  $I_{9}=t F_{rad}/(P_{th} r)=1/Bo$ ($Bo$: Boltzmann number). We recover the standard dimensionless numbers \cite{Mihalas99} which describe these radiating fluids.  The scaling laws are presented in table 2. 
   We see that for an ideal gas with an ionization conservation state, we have a single ($\delta_{5}$) parameter to rescale experiments, but if the ionization is not preserved, we have at least 
   three free parameters ($\delta_{5}$, $\delta_{9}$, $\delta_{14}$). Finally, we find that $\alpha$ (entropy) is conserved  for $\gamma=4/3$, which corresponds to a dominant photon regime. 
    Notice that if we set $P_{rad}=E_{rad}=0$, we find an extended scaling laws version of \cite{Murakami02}.
\begin{table}
\caption{ \small{\emph{Scaling laws of optically thick plasma (Column 1) and ideal gas (Column 2).}}}
\begin{center}
\begin{tabular}{llll}
\br
 \small{physical ratio} & \small{ratio} (scaling factor) & Ideal gas\\
\mr
$r/\tilde{r}$ & $a^{\delta_{14}+([n-5]/[4-\nu])\delta_{9}+([m+1/2]+\mu[n-5]/[4-\nu])\delta_{5}}$ & $a^{([m+1/2]+[n-5]/3)\delta_{5}}$  & \\
$t/\tilde{t}$& $a^{\delta_{14}+([n-7]/[4-\nu])\delta_{9}+(m+1+\mu[n-7]/[4-\nu])\delta_{5}}$  &  $a^{(m+1+[n-7]/3)\delta_{5}}$ & \\
$v/\tilde{v}$ & $a^{(2/[4-\nu])\delta_{9}+([4\mu+\nu-4]/[8-2\nu])\delta_{5}}$   & $a^{\delta_{5}/6}$  \\
$\rho/\tilde{\rho}$ &  $a^{\delta_{5}}$ & $a^{\delta_{5}}$  \\
$P_{th}/\tilde{P}_{th}$ & $a^{(4/[4-\nu])\delta_{9}+(4\mu/[4-\nu])\delta_{5}}$  & $a^{(4/3)\delta_{5}}$  \\
$T/\tilde{T}$ & $a^{(1/[4-\nu])\delta_{9}+(\mu/[4-\nu])\delta_{5}}$  & $a^{\delta_{5}/3}$ \\
$E_{rad}/\tilde{E}_{rad}$ & $a^{(4/[4-\nu])\delta_{9}+(4\mu/[4-\nu])\delta_{5}}$   & $a^{(4/3)\delta_{5}}$  \\
$F_{rad}/\tilde{F}_{rad}$ & $a^{(6/[4-\nu])\delta_{9}+([12\mu-4+\nu]/[8-2\nu])\delta_{5}}$   & $a^{(3/2)\delta_{5}}$  \\
$P_{rad}/\tilde{P}_{rad}$ & $a^{(4/[4-\nu])\delta_{9}+(4\mu/[4-\nu])\delta_{5}}$  & $a^{(4/3)\delta_{5}}$ \\
$\alpha/\tilde{\alpha}$ & $a^{(4/[4-\nu])\delta_{9}+([4\mu]/[4-\nu]-\gamma)\delta_{5}}$  & $a^{(4/3-\gamma)\delta_{5}}$ \\
$\mathcal{Q}_{rad}/\tilde{\mathcal{Q}}_{rad}$ & $a^{-\delta_{14}+([11-n]/[4-\nu])\delta_{9}+(\mu[11-n]/[4-\nu]-[m+1])\delta_{5}}$  & no source  \\
$\kappa_{0}/\tilde{\kappa}_{0}$ & $a^{\delta_{14}}$   & 1 \\
$\varepsilon_{0}/\tilde{\varepsilon}_{0}$ & $a^{\delta_{9}}$   & 1  \\
\br
\end{tabular}
\end{center}
\end{table}
 \section{Conclusion}
We presented the general scaling laws for two radiative regimes of major interest in laboratory and astrophysics situations. The number
of free parameters depends on the structure of the model: the more phenomena we add, the more difficult it is to rescale an experiment. 
However, requiring a \emph{partial} similarity ('\emph{almost}' equivalent regime) allows to add free parameters and study '\emph{almost}' astrophysical situations.

\end{document}